# Direct measurement of temporal correlations above the spin-glass transition by coherent resonant magnetic x-ray spectroscopy

Jingjin Song (1), Sheena K.K. Patel (1,2), Rupak Bhattacharya (1), Yi Yang (1), Sudip Pandey (1), Xiao M. Chen (3), M. Brian Maple (1), Eric E. Fullerton (2), Sujoy Roy (3), Claudio Mazzoli (4), Chandra M. Varma (5) and Sunil K. Sinha (1)

(1) Dept. of Physics, University of California San Diego, La Jolla, CA 92093-0319
(2) Center for Memory and Recording Research, University of California San Diego, La Jolla, CA 92093-0401
(3) Advanced Light Source, Lawrence Berkeley National Laboratory, Berkeley, CA 94720
(4) National Synchrotron Light Source II, Brookhaven National Laboratory, Upton, New York, 11973-5000
(5) Dept. of Physics, University of California Berkeley, Berkeley, CA 94720

**In the 1970's a new paradigm was introduced that interacting quenched systems, such as a spin-glass, have a phase transition in which long time memory of spatial patterns is realized without spatial correlations. The principal methods to study the spin-glass transition, besides some elaborate and elegant theoretical constructions, have been numerical computer simulations and neutron spin echo measurements . We show here that the dynamical correlations of the spin-glass transition are embedded in measurements of the four-spin correlations at very long times. This information is directly available in the temporal correlations of the intensity, which encode the spin-orientation memory, obtained by the technique of resonant magnetic x-ray photon correlation spectroscopy (RM- XPCS). We have implemented this method to observe and accurately characterize the critical slowing down of the spin orientation fluctuations in the classic metallic spin glass alloy Cu(Mn) over time scales of 1 to $10^3$ secs. Our method opens the way for studying phase transitions in systems such as spin ices, and quantum spin liquids, as well as the structural glass transition.**

The specific realization that a spin-glass encodes long-term memory of spin-orientations came through the introduction of an order parameter, by S.F. Edwards and P.W. Anderson, for the problem of interacting magnetic impurities with spins $S_i$ located randomly at sites $i$ in a metal or an insulator. The inspiration for this were experiments in which a cusp was found in the temperature dependence of the magnetic susceptibility of such "spin-glasses" (SG) which become sharper as the frequency of measurements is reduced towards zero. [1-5]  To facilitate the discussion of the SG transition , let us define a quantity

$$q(t,\tau) = \frac{1}{N}\sum_i \left\langle S_i(t)\right\rangle_T \left\langle S_i(t+\tau)\right\rangle_T$$

(1)

where N is the total number of spins, the subscript $T$ is the thermal average (henceforth the subscript $T$ will be dropped) and the sum over the random location of spins $i$ effectively averages over randomness after the temporal correlations are evaluated. We refer to the average over time t of q(t, τ) as q(τ). The Edwards-Anderson (EA) order parameter [5,6] may be defined as

$$Q = Lt_{\tau \to \infty} \left\langle q(t, \tau) \right\rangle_t \qquad (2)$$

where the subscript t denotes the average over t. The important point is that the memory of the pattern of randomness in the frozen configuration is retained when the thermal average $< S_i >_T$ of a spin located at a site $i$ is taken and its overlap with the same object at a later time is considered and found to be non-zero. Only then is the average over the spins at their random locations taken. The time-scales of the spin fluctuations leading to the transition are very long, as in the ordinary glass transition, quite unlike those of critical spin-wave modes in periodic magnetic systems. The nature of the SG transition has been a long-standing and important problem since this class of materials began to be studied in the 1970's. In particular, the approach to the SG transition encoded in the correlation function q(τ) over long times is of great interest.

The theory of spin-glasses continues to this day through highly developed mathematical methods to address the aspects of calculating measurable properties in random systems and the likelihood of multiple nearly degenerate ground states with metastable barriers penetrable at finite temperatures. The multiple states possible in a spin-glass with even more multiple routes of connection between such states has found parallels with theories of protein folding, prebiotic evolution, neural networks, stochastic linear programming problems, machine learning and other problems in computational sciences. [7-11]

While the SG problem has been mathematically fecund, very few experimental methods have given information on the microscopic correlation functions either in the spin-glass phase or in the transition to it. Macroscopic experiments noted above have been of course very instructive. There were early inelastic neutron scattering studies of spin-glasses which studied the two spin-correlations as the SG transition was approached, similar to the study of critical spin-fluctuations near the transition to ordered states in ferromagnets or anti-ferromagnets, but these did not lead to a clear indication of a sharp transition accompanied by critical phenomena [12-14]. However, the time scales and the nature of fluctuations in SGs are radically different from the typical time scales probed with neutron scattering (even for later, more detailed measurements with the neutron spin-echo (NSE) technique [15,16]) and since the order parameter of spin-glasses is itself a correlation of a pair of spins, only a specific four spin correlation can give information on the evolution in time of the fluctuations of the actual order parameter above the SG ordering temperature $T_g$.

Here, we show that the temporal correlations in the intensity of Resonant Magnetic X-ray Photon Correlation Spectroscopy (RM-XPCS), can be used to study the dynamical critical behavior of a Cu(Mn) SG by directly probing the requisite 4-spin correlation functions, which contain as the principal contribution, the fluctuations of the Edwards-Anderson (EA) order parameter. These are measured in real time on time scales much longer than those available to NSE or any inelastic scattering technique. The results show slow fluctuations and an unexpected slow decay of the fluctuation relaxation time with temperature, so that these fluctuations are still evident above 1.5 $T_g$, far above the critical regime of ordinary continuous transitions.

## XPCS Measurements and Results

The measurements were carried out on a $Cu_{0.88}Mn_{0.12}$ alloy film. Experimental details are given in the Methods section. Low field (100 Oe) dc magnetic susceptibility measurements as a function of temperature (Fig. 1) showed typical SG behavior for both field-cooled and zero-field-cooled measurements, with an estimated $T_g$ of 45 K.

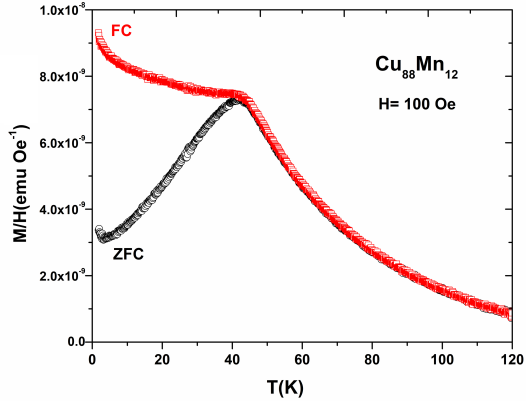

**Fig. 1**. Measured field-cooled (red) and zero-field-cooled (black) dc susceptibilities of the $Cu_{0.88}Mn_{0.12}$ sample as a function of temperature.

Resonant magnetic scattering (usually with the X-ray photons tuned to the L- or M-edge of the magnetic atoms) gives a large enhancement of the scattering amplitude due to the magnetic moment at resonance. If the photon energy is off-resonance, the scattering is only from charge scattering. The recorded signal on the detector consisted of speckle patterns [19-22] consisting of intensity fluctuations arising from the random phase interferences of the scattering from all the ions from the coherent X-ray beam. These consist of both small-angle charge scattering (which is static on the time scales of these measurements), and resonant magnetic scattering from the disordered and fluctuating Mn spins.

In the XPCS technique, the time-averaged intensity-intensity autocorrelation function is measured and normalized by the averaged intensity squared, resulting in a second-order correlation function $g_2(\mathbf{q},\tau)$. This is given by [19-22]

$$g_2(\mathbf{q},\tau) = \frac{\left\langle I_t(\mathbf{q},t)I_t(\mathbf{q},t+\tau)\right\rangle_t}{\left\langle I_t(\mathbf{q},t)\right\rangle_t^2}$$

(3)

where $<\ldots>_t$ represents an average over t. Here $I_t(\mathbf{q},t)$ represents the total (charge + magnetic) scattered intensity at the wave vector transfer $\mathbf{q}$.

The magnetic signal is not large even in the vicinity of the resonance X-ray energy, because of the low concentration of Mn ions, and the charge scattering dominates at small values of the wave vector transfer $\mathbf{q}$, but decreases rapidly with q. By comparing the resonant and off-resonant scattering we determined that in our sample the magnetic scattering dominates for q values >

0.005 Å$^{-1}$. Since the charge scattering and the magnetic dipole scattering are polarized perpendicular to each other, they do not interfere with one another, so that it is not possible to heterodyne off the (static) charge scattering. Instead the latter acts as a static background under the signal, so the RM-XPCS analysis has to be slightly modified, as discussed below. We have

$$I_t(\mathbf{q},t) = I_c(\mathbf{q},t) + I_m(\mathbf{q},t) \qquad (4)$$

where $I_c$ is the intensity of the charge scattering, $I_m$ that of the magnetic scattering and $I_t$ is the total scattered intensity (all at measurement time t). In the dipolar approximation for quasi-elastic resonant magnetic scattering [17,18] $I_m(\mathbf{q},t)$ is given by

$$I_m(\mathbf{q},t) = \frac{1}{N}C\sum_{ij}\left\langle(\hat{\mathbf{e}}_{in}\times\hat{\mathbf{e}}_{out}).\mathbf{S}_i(t)(\hat{\mathbf{e}}_{in}\times\hat{\mathbf{e}}_{out}).\mathbf{S}_j(t)\right\rangle e^{-i\mathbf{q}.(\mathbf{R}_i-\mathbf{R}_j)} \qquad `1(5)$$

if we keep only the dominant linear term in the magnetic spins in the X-ray magnetic scattering length. In Eq. (5), <..> stands for thermal average at time t; the sum over sites carries out the configuration average; $\hat{\mathbf{e}}_{in}$ and $\hat{\mathbf{e}}_{out}$ represent unit polarization vectors for the incident and scattered photons on the sample respectively, $\mathbf{S}_i$ and $\mathbf{S}_j$ are the spins on lattice sites $\mathbf{R}_i$ and $\mathbf{R}_j$ respectively, and the spin operators are equal time operators whose thermal averages are evaluated at time t, since there is no energy selection of the scattered X-ray beam. (In general, the actual time differences for the 2 spin operators can vary over times of the order of the inverse of the energy resolution of the experiment. In our case, if the total measurement time is divided into many short time intervals, as it is in XPCS experiments, it actually represents the equal time average over a particular measuring interval). Finally, N is the total number of spins in the measuring volume and C is a combination of instrumental and geometrical factors and resonant dipole matrix elements. It can easily be shown, that if the scattered beam makes a small angle θ to the incident beam direction (small angle approximation),

$$(\hat{\mathbf{e}}_{in}\times\hat{\mathbf{e}}_{out}).\mathbf{S}_i \cong S_i^z \qquad (6)$$

where $S_i^z$ is the component of $\mathbf{S}_i$ along the incident beam direction. We neglect corrections of the order θ, which is equivalent to setting $\mathbf{q}$ =0 in Eq. (5). Thus, we can write Eq. (5) as

$$I_m(\mathbf{q},t) = \frac{1}{N}C\sum_{ij}\left\langle S_i^z(t)\right\rangle\left\langle S_j^z(t)\right\rangle \qquad (7)$$

where the brackets represent thermal averages of the spin operators at the time t. In our experiment, we found that the above correlation function, averaged over long times t, indeed showed very little q-dependence at the small q-values over which the measurements were done, consistent with the above approximation. This is also consistent with the normal assumptions made about "ideal" spin glasses. Then the autocorrelation function for the magnetically scattered intensities can be approximated by

$$\left\langle I_m(\mathbf{q},t)I_m(\mathbf{q},t+\tau)\right\rangle_t = \frac{1}{N^2}C^2\sum_{ij}\left\langle\left\langle S_i^z(t)\right\rangle\left\langle S_j^z(t)\right\rangle\left\langle S_i^z(t+\tau)\right\rangle\left\langle S_j^z(t+\tau)\right\rangle\right\rangle_t$$

$$+\frac{1}{N^2}C^2\sum_{ij}\left\langle\left\langle S_i^z(t)\right\rangle\left\langle S_j^z(t)\right\rangle\right\rangle_t\sum_{kl\neq ij}\left\langle\left\langle S_k^z(t+\tau)\right\rangle\left\langle S_l^z(t+\tau)\right\rangle\right\rangle_t$$

$$(8)$$

In the first term, we have kept the autocorrelation over t for the same pairs of spins at times t and t+τ , while in the second term we have decoupled the time averages over different pairs of spins as if they were statistically uncorrelated.  By Eq. (7), the second term is equal to <[I_m(**q**,t)]>_t² to order (1/N). Thus, after rearranging some factors, we may write Eq. (8) as

$$\left\langle I_m(\mathbf{q},t)I_m(\mathbf{q},t+\tau)\right\rangle_t = \chi_{SG}(\tau) + \left\langle I_m(\mathbf{q},t)\right\rangle_t^2 \qquad (9)$$

where

$$\chi_{SG}(\tau) = \frac{1}{N^2}C^2\left\langle\sum_i\left\langle S_i^z(t)\right\rangle\left\langle S_i^z(t+\tau)\right\rangle\sum_j\left\langle S_j^z(t)\right\rangle\left\langle S_j^z(t+\tau)\right\rangle\right\rangle_t = \frac{1}{N^2}C^2\left\langle q(t,\tau)^2\right\rangle_t \qquad (10)$$

In principle, $\chi_{SG}$ could be **q**-dependent, but based on our current data, we can neglect the **q**-dependence. In our results, $\chi_{SG}$ can be related to the fluctuations of the EA order parameter (see Eq. (1)). In fact, below $T_g$, $\chi_{SG}$ becomes independent of τ, and proportional to $Q^2$ as defined in Eq. (2).  The inner brackets represent instantaneous thermal averages, while the big bracket represents an average over t. We note that the factor C contains the solid angle ΔΩ subtended by the group of pixels contributing to $I_m(\mathbf{q},t)$ so that the quantity above is of order $(\Delta\Omega/4\pi)^2$.

In XPCS, the total intensity autocorrelation function $g_2(\mathbf{q},\tau)$, can be written as

$$g_2(\mathbf{q},\tau) = 1 + \beta\frac{\left\langle I_t(\mathbf{q},t)I_t(\mathbf{q},t+\tau)\right\rangle_t - \left\langle I_t(\mathbf{q},t)\right\rangle_t^2}{\left\langle I_t(\mathbf{q},t)\right\rangle^2} \qquad (11)$$

where β is the contrast factor [19-22] arising from the partial coherence of the X-ray beam. Using Eq. (4), and the fact that the charge scattering is independent of time, and that there is no interference between charge and magnetic scattering, we obtain

$$g_2(\mathbf{q},\tau) = 1 + \beta\frac{\left\langle I_m(\mathbf{q},t)I_m(\mathbf{q},t+\tau)\right\rangle_t - \left\langle I_m(\mathbf{q},t)\right\rangle_t^2}{\left\langle I_t(\mathbf{q},t)\right\rangle_t^2} \qquad (12)$$

which by Eq. (9) can be written as

$$g_2(\mathbf{q},\tau) = 1 + \beta\frac{\chi_{SG}(\tau)}{\left\langle I_t(\mathbf{q},t)\right\rangle_t^2} \qquad (13)$$

Thus, the function [$g_2(\mathbf{q},\tau) - 1$] measures the 4-spin correlations as given in Eq. (10) above.

We may summarize our experimental results and consistency checks for measurements of $g_2(\mathbf{q},\tau)$ as follows (For details see Methods section):

(a) First, we verified that $g_2$ was independent of the starting time $t_0$ for measuring the time-dependent intensity autocorrelation function.

(b) Then, in order to take into account the possible intensity fluctuations of the incident beam, the $g_2$ functions were corrected by dividing by the time autocorrelation function of the incident beam intensity.

(b) The off-resonance measurements (pure charge scattering) at all q-values, and the on-resonance measurements of $g_2(\mathbf{q},\tau)$ at small q (where charge scattering dominated) showed no time dependence, as expected. The larger q ($> \sim 0.005$ Å$^{-1}$) where the spin scattering dominated showed time dependence , but there was very little dependence of the magnetic scattering contribution to $g_2(\mathbf{q},\tau)$ on $\mathbf{q}$, over the range of wavevectors studied, as already discussed above.

(c) The measured time dependence of $[g_2(\mathbf{q},\tau)-1]$ could be well fitted by a simple exponential. (as shown in Fig. 2 for the measurements at 74 K).

(d) The contrast factor $\beta$ as measured by Eq. (11) from the un-normalized curves for $[g_2(\mathbf{q},\tau)-1]$ came out to be of order $10^{-3}$, typically 2 orders of magnitude smaller than in a conventional XPCS experiment. This can be attributed, in part, to the weakness of the magnetic scattering signal, but most likely arises mainly from two causes: (1) the fact that there are many spin excitations which decay on much faster length scales and whose contributions have vanished before our first measuring time frame is completed (as seen in the NSE experiments [15,16], for example) and (2) because we are left with only same-spin correlation functions in the final time-averaged $g_2$ function which reduces the intensity by a factor $1/N$ compared to that from normal correlation functions which involve interference scattering  from different spins. The situation is somewhat reminiscent of incoherent vs. coherent neutron scattering from assemblies of atoms.

The principal results are given below.

The time-dependence of the $g_2$ functions can be fitted very well by the form

$$g_2(q,t) = C_1 + C_2 e^{-(t/\tau_0)} \qquad (14)$$

where the relaxation time $\tau_0(T)$ shows little dependence on q-values but is temperature-dependent.

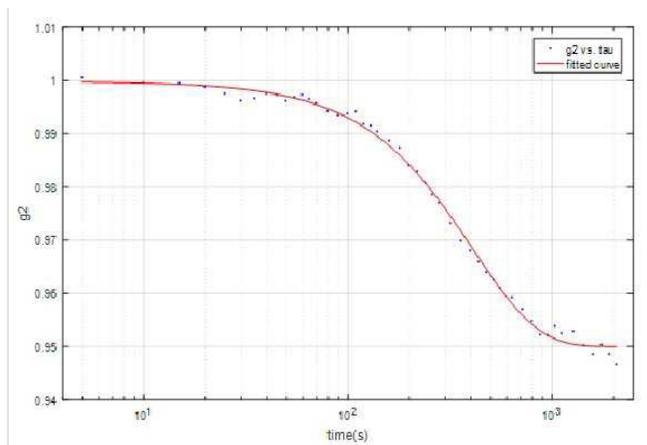

**Fig. 2.** Experimental data and fitted exponential curve for the function [$g_2$(q,t)-1] and q = 6.4 x $10^{-3}$ Å$^{-1}$ at T= 74 K. (curve has been normalized to 1 at t = 0)

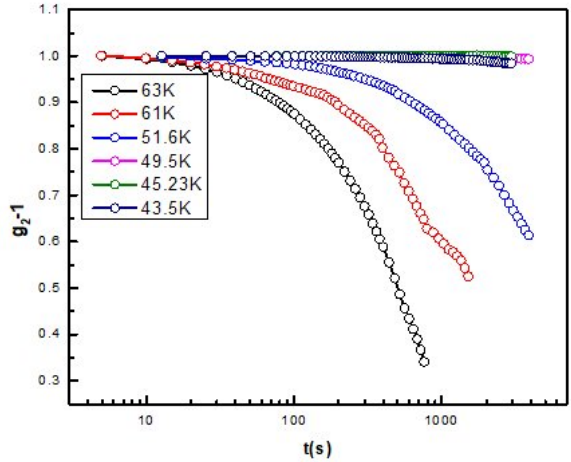

**Fig. 3.** The functions [$g_2$(q,t)-1] at q =6.4 x $10^{-3}$ Å$^{-1}$ for several different temperatures. These curves have all been normalized to unity at t = 0.

Fig. 3 shows how the relaxation time increases rapidly as the temperature is cooled towards $T_g$. However, there is a limiting time of ∼ 3600 secs, because of the decay in time of the autocorrelation function of the incident beam intensity from the synchrotron itself. Although this was taken into account to some extent by normalizing the $g_2$ functions by those of the incident beam, it limited our effective measurements of the spin dynamics to time scales of < 2x10$^4$ secs.

The measured temperature dependence is shown in Fig. 4(a), and could be fitted reasonably well with the form

$$\tau_0(T) = \frac{A}{(T - T_g)^B}$$

(15)

The fit yielded values of 44.12 K for $T_g$ and B = 2.68. While the fitted value is quite close to the value of $T_g$ from susceptibility measurements, the value of B appears to indicate that the dynamical critical fluctuations we observe decay much more slowly than expected from the dynamical critical exponent (zv) observed in earlier computer simulations [23] and nonlinear susceptibility measurements to be ∼ 7 [24]. It should be noted however that those simulations were for a nearest-neighbor random ± J exchange Ising model on a cubic lattice. On the other hand, the mean field value of (zv) is 2 and values even lower have been quoted experimentally [25,26].

We also attempted to fit the temperature dependence of the relaxation with the Vogel-Fulcher form

$$\tau_0(T) = \tau_1 \exp(E_0 / T - T_0)$$

(16)

familiar in the discussion of the glass transition [27,28]. This is shown in Fig.4(b). The fit is equally good, but the corresponding value of $T_0$ is 36.51 K which is a few degrees below the experimentally determined value of $T_g$.

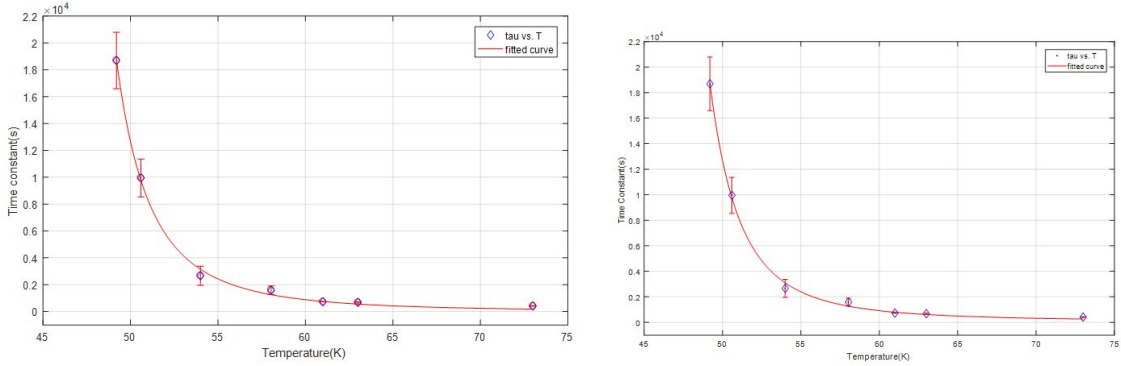

**Fig.4.** (a) Temperature dependence of the relaxation time vs. temperature with power law fit. (b) Temperature dependence of the relaxation time vs. temperature with Vogel-Fulcher fit.

These measurements constitute the first direct measurement of the temperature dependence of the time-dependent susceptibility related to the SG (EA) order parameter fluctuations, and show slowing down of the fluctuations as $T_g$ is approached from above, consistent with power-law behavior. The power law is inconsistent with the dynamical critical exponent obtained from computer simulations [23], and those deduced indirectly from measurements of the non-linear magnetic susceptibility [24], although a wide range of values is found in the literature [25,26]. Note that in the notation of critical phenomena, where we define the reduced temperature $t_r = (T - T_g)/T_g$, our range of dynamical measurements covers the range from $t_r = 0.01$ to $t_r = 0.78$. It is thus possible that they do not measure within the true critical regime. Finally, we note that the accuracy and number of temperature points of the present measurements is likely not high enough to yield a reliable value of the dynamical critical exponent, but these measurements do represent an interesting new technique to measure slow fluctuations in frustrated magnetic systems, and indicate the need for further detailed measurements on similar systems.

It should be noted that Werner and collaborators, in a series of papers, have shown, using neutron scattering, that CuMn alloys in the same concentration ranges as the samples studied here, show pronounced diffuse scattering peaks at points in reciprocal space, corresponding to short-range spin-density-wave (SDW) order [29-31]. This order presumably arises from Mn-rich clusters where the RKKY interactions can induce local order. It has been claimed by several authors [32-34] that spin glass order can coexist with even long-range antiferromagnetic order. Further, our results focus on fluctuations around q = 0, quite separate from the q-regions where the short-range correlations of the SDW state are found. Therefore we believe that our measurements are indeed related to the critical SG fluctuations of the CuMn alloy studied.

Interestingly, the present technique could also be potentially useful in studying the fluctuations of entangled singlet spin pairs that have been proposed for the case of the Resonating valence bond state in quantum spin liquids [35,36], or the slow fluctuations in spin ices [37,38], since these would also involve 4-spin correlation functions. Although the present method concentrates on very slow fluctuations, much faster fluctuations can also be studied by the RM-XPCS technique at time scales of nano seconds or less by using the delayed 2-pulse speckle contrast measurement methods that have recently been developed for free-electron X-ray laser sources [39-41].

## References

[1] J.A. Mydosh, Rep. Prog. Phys. **78** 052501 (2015)

[2] J.A. Mydosh, Journal of Magnetism and Magnetic Materials **7** 237 (1978)

[3] C.Y.Huang, Journal of Magnetism and Magnetic Materials **51** 1 (1985)

[4] *Spin Glasses and Random Fields*, A.P.Young, Ed., Series on Directions in Condensed Matter Physics, Vol. 12, World Scientific Singapore, 1998

[5] P.W.Anderson, Physics Today **41**,1,9 (1988); Physics Today **42**, 9 (1988); Physics Today **43**, 9 (1988).

[6] S.F. Edwards and P.W. Anderson, J.Phys. **F**: Metal Physics **5**, 965 (1975)

[7] J.D.Bryngelson and P.G.Wolynes, Proc. Nat. Acad. Sciences **84**, 7524 (1987)

[8] M.Mezard, G. Parisi and M.A.Virasoro, in "Spin Glass Theory and Beyond" (World Scientific Lecture Notes in Physics, Vol. **9**, World Scientific Press, Singapore 1986)

[9] V.Dotsenko, in "*An Introduction to the Theory of Spin Glasses and Neural Networks*" (World Scientific Lecture Notes in Physics, Vol. **54**, World Scientific Press, Singapore 1995)

[10] S. Franz, Phys. Rev. Lett. **87**, 127209 (2001)

[11] R. Monasson, J. Phys. A: Math. Gen., **31**,513 (1998)

[12] A.P.Murani and A. Heidemann, Phys. Rev. Lett. **41**, 1402 (1978)

[13] F. Mezei, J.Mag. Mag. Mat. **31-34**, 1327 (1983)

[14] F.Mezei and A.P.Murani, J.Mag. Mag. Mat. **14**, 211 (1979)

[15] C. Pappas et al., Phys. Rev. **B 68**, 054431 (2003)

[16] R.M.Pickup et al., Phys. Rev. Lett. **102**, 097202 (2009)

[17] J.P.Hannon et al., Phys. Rev. Lett. **61**, 1245 (1989)

[18] M.Blume, J.Appl. Phys. **57**, 3615 (1985)

[19] Alec R. Sandy, Qingteng Zhang,and Laurence B. Lurio, Ann. Rev. Mater. Res. **48**,167 (2018)

[20] G. Grubel, A.Madsen, and A. Robert, "X-ray Photon Correlation Spectrocopy" in *Soft Matter Characterization* (Eds.R. Borsali and R. Pecora) Springer-Verlag , Chapter 18,2008.

[21] M.Sutton, C. R. Physique **9**, 657 (2008)

[22] Sunil K. Sinha, Zhang Jiang, and Laurence B. Lurio, Adv. Mater. **26**, 7764 (2014)

[23] A.T.Ogielski, Phys. Rev. **B 32**, 7384 (1985)

[24] L.P.Levy, Phys. Rev. **B 38**, 4963 (1988)

[25] G. Parisi et al., J. Phys. **A**, Math. and Gen., **30**, 7115 (1997)

[26] M.K.Singh et al.,Phys. Rev **B 77**, 144403 (2008)

[27] J. Langer, Physics Today **60**, 8 (2007)

[28] J. C. Dyre, Rev. Mod. Phys. **78**, 953 (2006)

[29] J.W.Cable et al., Phys. Rev.**B 29**,1268 (1984)

[30] F.J.Lamelas et al., Phys. Rev.**B 51**,621 (1995)
[31] S.A. Werner, Comments in Cond. Matter Phys. **15**, 55 (1990)
[32] P.-Z. Wong et al., Phys. Rev. Lett. 55, 2043 (1985)
[33] S. Chillal et al., Phys. Rev. **B 87**, 220403 (R) (2013)
[34] V.H.Tran et al., J. Phys.: Condens. Matt. 17, 3597 (2005)
[35] P.W. Anderson, Mat. Res. Bull. **8**, 153 (1973).
[36] Y. Shimizu et al. Phys. Rev. Lett. **91**, 107001(2003)
[37] L.R.Yaraskavitch et al., Phys. Rev. B **85**, 020410 (R) (2012)
[38] K. Matsuhira,Y. Hinatsu and T. Sakakibara, J. Phys. Condens. Matter **13** L737 (2001)
[39] C. Gutt et al., Opt. Express **17**, 55 (2009)
[40] M.H. Seaberg et al., Phys. Rev. Lett. **119**, 067403 (2017)
[41] Y.W. Sun et al., Optics Letters **44**, 2582 (2019)

**Methods**

The sample was a polycrystalline film of the alloy $Cu_{0.88}Mn_{0.12}$ of thickness $\sim 400$ nm prepared by co-sputtering Cu and Mn in the proper ratios. This was deposited on a 7.5 mm x 7.5 mm SiN/Si substrate film with a 1 mm x 1 mm window of SiN of thickness 100 nm, so that a soft X-ray beam could be transmitted through it and through the sample in a transmission geometry scattering experiment. The sample was mounted in a He flow cryostat, initially on beamline 23-ID-1 (CSX) at the NSLS-II Light Source, and subsequently on beamline 12.0.2 at the Advanced Light Source (ALS). These beamlines have the capability of producing a coherent beam of photons by transmission through a 10-$\mu$m-diameter (5 $\mu$m at ALS) pinhole at photon energies tunable around the Mn $L_3$ edge at $\sim 641$ eV. Measurements were made at 636 eV, slightly below the resonant edge energy (to optimize the resonant magnetic dipole scattering [17,18], while mitigating the peak absorption at the resonance) and also at 10 eV below this energy to study the non-resonant or purely charge scattering. The incident photon beam was linearly polarized in the horizontal plane. Measurements were made in transmission in the forward direction in small angle geometry. The scattered photons were recorded on a 2D detector.

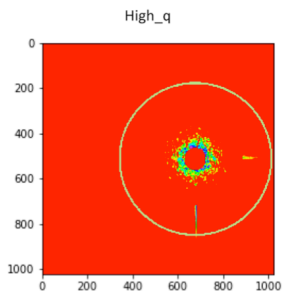

**Fig. 5.** Typical speckle pattern on the area detector at the ALS in a 5 second exposure at room temperature. Speckles due to intense static small angle charge scattering are seen around the beam stop positioned at **q** =0. Magnetic speckles (typically measured at the larger q-values

indicated by the ring at q = 6.4 x10⁻³ Å⁻¹, the so-called "high q" values) cannot be seen on this intensity scale, being too weak and rapidly fluctuating. Streaks seen in image are static artifacts from the sample.

**Examining the difference between resonant magnetic scattering and charge scattering in the g₂-functions.**

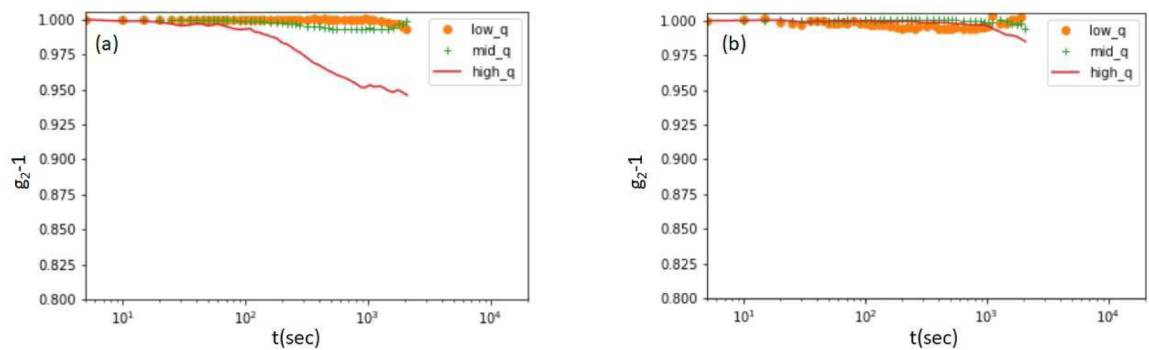

**Fig. 6.** Plots of $(g_2 - 1)$ vs. time at T = 80 K (a) on resonance at the Mn $L_3$ edge and (b) off-resonance. The q- values in both cases were: high q=6.4 x10⁻³ Å⁻¹, mid q =4.0 x 10⁻³ Å⁻¹, and low q=1.9 x 10⁻³ Å⁻¹. At low q-values and at off-resonance, predominantly charge scattering is observed, with no decay of $(g_2 - 1)$, whereas for larger q and on resonance, predominantly magnetic scattering is observed with concomitant decay of $(g_2 - 1)$.

**Test of independence of g₂ function on start times.**

By examining $g_2(q, t_2-t_1)$ for different starting times $t_1$ in specific runs, we verified that it was independent of the starting time $t_1$ and depended only on the time difference, as expected. See figure below.

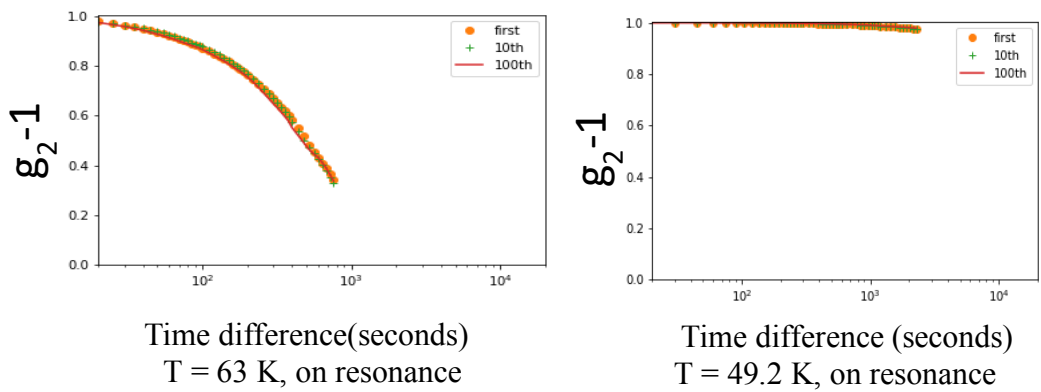

**Fig.7.** The function $(g_2 - 1)$ plotted vs. time difference between start and end times for starting

times at t =0, 50 seconds and 500 seconds respectively, for q = 6.4 x 10⁻³ Å⁻¹ and 2 different temperatures with the photon beam at the energy of the Mn $L_3$ resonance energy. The time frame for collecting the time-dependent data was 5 seconds.

**Dependence of ($g_2$ −1) on q**

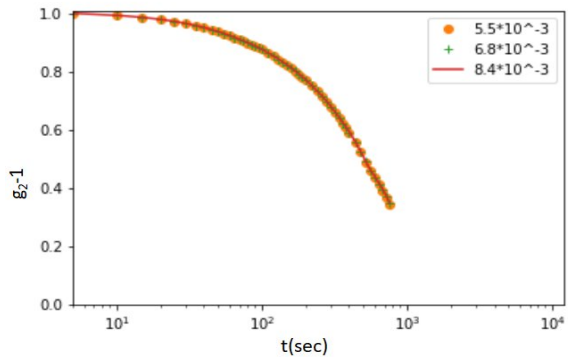

**Fig. 8.** ($g_2$ −1) plotted for several q-values in the region where magnetic scattering as dominant. The curves superimpose, showing no q-dependence at these q-values.

**Corrections for decay of incident beam.**

The final $g_2$-functions were divided by the $g_2$-function (self auto-correlation function) of the incident beam itself to correct for its decay with time. This procedure is based on the fact that the intensity fluctuations of the incident beam and the scattering cross-section are statistically independent. Fig. 8 illustrates the $g_2$-function of the intensity of the main beam, taken as the time-dependent integrated intensity in the 2D detector, as a function of time difference.

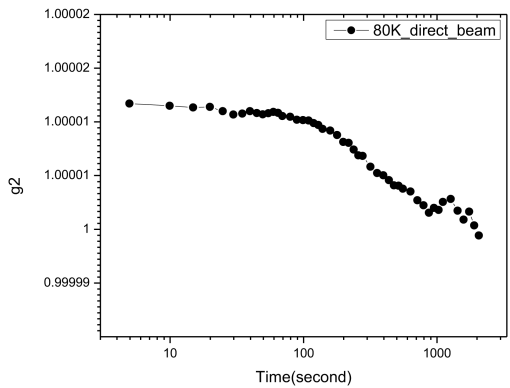

**Fig. 9.** Intensity-intensity autocorrelations of the incident beam as a function of time difference. Intensity was taken as proportional to the total counts in the 2D detector.

## Data availability

The data that support the findings of this study are available from the corresponding author upon reasonable request.

## Acknowledgements

We wish to acknowledge the help from the staff of the CSX beamline at NSLS II, and staff members at ALS in the experiments at the 12.0.2 coherent beamline at ALS. We thank Kalyan Sasmal and Sheng Ran for assistance with magnetic measurements made at UCSD. The research was supported by grant no. DE-SC0003678 from the Division of Basic Energy Sciences of the Office of Science of the U.S. Dept. of Energy. MBM acknowledges the support of the US Department of Energy, Office of Basic Energy Sciences, Division of Materials Sciences and Engineering, under Grant No. DE-FG02-04ER46105. This research used beamline 23-ID-1of the National Synchrotron Light Source II, a U.S. Department of Energy (DOE) Office of Science User Facility operated for the DOE Office of Science by Brookhaven National Laboratory under Contract No. DE-SC0012704. Work at the ALS, LBNL was supported by the Director, Office of Science, Office of Basic Energy Sciences, of the US Department of Energy (Contract No. DE-AC02-05CH11231).

**Author Contributions:** JS, RB and SKKP contributed equally to this work. SR and CM prepared the beamlines at ALS and NSLS respectively for the experiments and helped to collect the data. SKKP and EEF prepared the samples, MBM made the magnetic susceptibility measurements, JS, RB, YY, SP, XMC, and SKS collected and analyzed the data at the synchrotron sources, CMV and SKS developed the formalism for analyzing the XPCS data, and SKS conceived of and planned the experiments. MBM,EEF, CM, SR, CMV and SKS discussed the final form of the paper.

## Competing interests
The authors declare no competing interests.